\documentclass[a4paper]{jpconf}
\usepackage{amssymb}
\usepackage{amsthm}
\usepackage{latexsym}
\newcommand\beq{\begin{equation}}
\newcommand\eeq{\end{equation}}
\newcommand\bea{\begin{eqnarray}}
\newcommand\eea {\end{eqnarray}}

\def\un\a{{\underline\alpha}}

\def\a{{\alpha}}

\begin{document}

\title{\textbf{The Emergence of Probabilities in\\ Anhomomorphic
Logic}\footnote{Based on talk given by P. Wallden at the DICE 2008
conference}}

\author{\textbf{Yousef Ghazi-Tabatabai}}

\address{Blackett Laboratory, Imperial College,\\
        London, SW7 2AZ, U.K.}

\ead{\textbf{yousef.ghazi05@imperial.ac.uk}}

\author{\textbf{Petros Wallden}}

\address{Raman Research Institute, Theoretical Physics Group \\ Sadashivanagar, Bangalore - 560
080, India}

\ead{\textbf{petros.wallden@gmail.com}}

\date{}

 \vspace{1cm}

\begin{abstract}

\noindent Anhomomorphic logic is a new interpretation of Quantum Theory (due
to R. Sorkin). It is a \emph{histories} formulation (c.f. consistent
histories, quantum measure theory). In this approach, reality is a
co-event, which is essentially an assignment of a truth value
\{True, False\} to each question. The way this assignment is done
mimics classical physics in as much as possible, allowing however
for sufficient flexibility to accommodate quantum `paradoxes', as is
shown by the analysis of Kochen-Specker theorem. In this
contribution, after briefly reviewing the approach, we will examine
how probabilistic predictions can arise. The Cournot principle and
the use of approximate preclusions will play a crucial role. Facing
similar problems in interpreting probability as in classical
probability theory, we will resort to the \emph{weak} form of
Cournot principle, where possible realities will be preclusive
co-events and the quantum measure is used to obtain predictions.
Examples considered, includes the fair coin and the double slit
pattern arguably one of the most important paradigms for quantum
theory.
\end{abstract}

\section{Introduction}

Quantum theory challenges the picture we had in classical physics
about what reality is. In order to retain the classical picture
alternative formulations, such as hidden variables, were proposed,
facing however serious problems due to the severe restrictions
from observations that result from Bell's
inequalities \cite{Clauser:1969ny} and the Kochen-Specker theorem
\cite{KS theorem}. Quantum theory (as it stands now) requires some
external (to the system) observer, in order to ``make sense''.
However, for the needs of quantum cosmology for example, where we
have a truly closed system, we do lack an interpretation. Consistent
histories \cite{Gri84,Omn88a,GH90b}, can be seen as an attempt to deal with this, that also removed in some way the very
special role that time has in standard quantum theory. The latter is
another desired feature, if one is interested in building a quantum
theory of gravity where time is in same footing with space.
Consistent histories however, failed to provide a fully satisfactory
interpretation of the quantum theory of closed systems due to the
context dependence ( or else the dependence of predictions on the
consistent set realized, which arises due to the fact that there
exist many incompatible consistent sets). As a development came a
novel interpretation which retains essentially the same mathematical
structure (history space and decoherence functional) the so-called,
\emph{anhomomorphic logic} \footnote{Also known with different names
as for example ``co-event interpretation''.}
\cite{Sorkin:2006wq,Sorkin:2007uc}. In the next section we introduce
this approach briefly, before we come back to the core of this
contribution, which is how probabilistic predictions are dealt
within this approach\footnote{More details can be found in Ref.
\cite{Emerent Probabilities}.}.

\section{Introducing Anhomomorphic Logic}
\subsection{Classical Physics, Histories and Logic}

Let us first consider classical physics, in which we
use Histories and (classical) Logic in order to ask questions about
nature.

We have a set $\Omega$ (call it \emph{History Space}) of all
possible histories $h_i$. We can think of each $h_i$ as a trajectory
(or more generally a full specification of the particle in every
moment of time). In classical physics one and only one of these
histories is actually realized. However, in stochastic physics
(non-deterministic) we do not know which one is realized, but have a
probability measure on $\Omega$\footnote{Histories that have measure
zero, are never realized, and thus are not even possible realities.
This will be of use later.}. All possible questions about a system
can now be rephrased in terms of asking whether the real history
belongs to some subset of $\Omega$. So for example if we want to ask
whether the particle is at the interval $\Delta$ at time $t$, we ask
if the realized history $h$ belongs to the subset
$\Delta_t:=\{h_i|h_i(t)\in \Delta\}\subset\Omega$.

Associated with $\Omega$ is its power set $\mathcal U$ (set of
subsets of $\Omega$), that has a Boolean algebra structure with intersection as multiplication and symmetric difference ($A\triangle B :=(A\cup
B)\setminus(A\cap B)$) as addition.

We also have  the set of truth values $\mathcal T$ ( e.g. \{True,
False\}) which also has a Boolean algebra structure (that of $\mathbb{Z}_2$, identifying `True' with $1$ and `False' with $0$). Finally we have
the possible valuations $\phi_i$. A valuation, $\phi_i$ is an
assignment of a truth value to each question, i.e. in other words a
map from $\mathcal U$ to $\mathcal T$. We moreover require that this
valuation respects the Boolean structures of $\mathcal U$ and
$\mathcal T$ by being a homomorphism:

\bea \phi(A\triangle B)&=&\phi(A)+\phi(B)\nonumber\\ \phi(A\cap
B)&=&\phi(A)\phi(B)\eea It can be shown that if one requires the
maps $\phi$ to be homomorphic there is a one to one correspondence
between these homomorphic maps and single histories, i.e. each
homomorphism corresponds to a characteristic map of a
history:\beq\phi_i | \phi_i(A)=1 \textrm{ iff } h_i\in A\eeq

\subsection{Consistent Histories and Quantum Measure}

In quantum theory the above picture does not hold. The only fully
developed histories formulation (at present) is the consistent
histories approach (also known as decoherent histories) Ref.
\cite{Gri84,Omn88a,GH90b}\footnote{Mathematically is very close to
the well known sum-over-histories formalism of Feynman, but the
interpretation is quite different}. The basic elements are histories
of the system and to each pair of histories a complex number is
assigned by the means of the \emph{decoherence functional} and it
corresponds to the interference between these histories. It is
defined as:

\bea D(A,B)&=&D^*(B,A)\nonumber\\
D(A\sqcup B,C)&=&D(A,C)+D(B,C)\nonumber\\
D(A,A)&\geq& 0\nonumber\\
\sum_{h,h'}D(h,h')&=&1 \eea The diagonal elements correspond to
\emph{candidate} probabilities and under certain
circumstances\footnote{If it belongs to a partition that the
elements pairwise have zero interference.} we can interpret this (real)
number as the probability of this history (or in general coarse
grained history i.e. subset of histories) actually occuring (this
probability is conditional on the classical domain, or consistent set, that is actually realized). For more details the reader is
referred to the original references \cite{Gri84,Omn88a,GH90b}.

This candidate probability interpretation led R. Sorkin to think of the diagnoal elements of a decoherence functional as a \emph{quantum measure}, as they mimic some of the properties of a classical measure, and can be thought of as the first step in a chain of generalizations of classical measure theory (see Refs. \cite{qmeasure1,qmeasure2}). We will use the notation
$\mu(A):=D(A,A)$ for the quantum measure. While it is normalized to
unity and is positive-definite, it fails to obey the ``additivity of
disjoint regions of the sample space'', a necessary requirement to
have a probability measure, \beq\mu(A\sqcup
B)\neq\mu(A)+\mu(B)\eeq This is due to interference (c.f.
double slit). Note though that the quantum measure \emph{does} obey the weaker

\beq\label{quantum measure}\mu(A\sqcup B\sqcup C)=\mu(A\sqcup
B)+\mu(A\sqcup C)+\mu(B\sqcup C)-\mu(A)-\mu(B)-\mu(C)\eeq Which
means that in quantum theory, there is no fundamentally new three
paths interference. In particular, it can be shown that any quantum
measure that obeys Eq. (\ref{quantum measure}) can arise from some
decoherence functional (as defined above, and thus the equivalence
between quantum measures and decoherence functionals.).

Due to the difference of the quantum measure compared to classical
measure, the picture of classical histories and logic analyzed in
the previous section, cannot be retained. The most striking problem,
comes essentially from the Kochen-Specker theorem, and is discussed
in Ref. \cite{Dowker:2007ma}. A simplified version
form of this problem comes from the three slit experiment, where we
have $\mu(A\sqcup B)=\mu(B\sqcup C)=0$ but $\mu(A\sqcup C)\neq 0$.
The first two imply that all the set $\{A,B,C\}$ is impossible
(since it is covered by measure zero sets), however one subset, $\{A,C\}$, \emph{is} possible.

\subsection{Anhomomorphic Logic}

We can see now that in order to be compatible with quantum theory,
one needs to alter something in the classical picture. Anhomomrphic
logic is the approach where we maintain the same set of possible
questions (namely $\mathcal U$), same possible truth values
($\mathcal{T}= \{\textrm{True, False}\}$) but we change the allowed
valuation maps, $\phi_i$, by weakening the requirement the map is a
homomorphism. These maps (no longer homomorphic) are called
\emph{co-events}. This approach was initiated by R. Sorkin in Refs.
\cite{Sorkin:2006wq,Sorkin:2007uc}. As already stated, F. Dowker and
Y. Ghazi-Tabatabai in Ref. \cite{Dowker:2007ma} showed how the
suggested approach evades
 Kochen-Specker theorem\footnote{See also Ref. \cite{Quantum Covers} for related work.} and thus is a good candidate for a realistic
 interpretation of quantum theory.

While we want to weaken the requirement to be homomorphism we need
it to maintain sufficient structure to be able to make deductions
(deductive logic), retain some sense of reality and in the same time
be able to accommodate the paradoxes of quantum theory. There are
three conditions we require.

\begin{enumerate}
\item We retain the preservation of multiplication under $\phi$ but no longer require the preservation addition in general,

\bea \phi(A\cap B)&=&\phi(A)\phi(B)\nonumber\\ \phi(A\triangle
B)&\neq&\phi(A)+\phi(B)\eea This is called \emph{multiplicative}
co-event\footnote{Earlier alternative definitions were use as in
Ref. \cite{Sorkin:2006wq} but this turned out to be the most
satisfactory due to several reasons.}.   It has the following
desirable properties.

\begin{itemize}
\item[(a)] \beq \phi(A)=1\textrm{ and }A\subset
B\Rightarrow\phi(B)=1\eeq This is the basic inference rule (``modus
ponens''), and can be used to make deductive proofs\footnote{e.g.
``I am physicist''-True, along with ``Physicists are humans''-True,
implies that ``I am human''-True. }. However we \emph{cannot} use
proofs by contradiction (excluded middle), since

\beq\phi(A)=0 \nRightarrow\phi(\Omega\setminus A)=1\eeq This is the
case in intuitionistic logic also (see also constructive mathematics
for mathematics using only deductive proofs). Note though, that the
following \emph{does} hold:

\beq\phi(A)=1 \Rightarrow\phi(\Omega\setminus A)=0\eeq
\item[(b)] It can be shown, that each multiplicative co-event\footnote{at least for finite dimensional history space $\Omega$ but similar considerations can be generalized after taking
care.} $\phi_A$ corresponds to a subset $A$ called \emph{dual} of
the co-event\footnote{In literature, the dual is also referred
occasionally as the ``suport''.}, such that

\beq\phi_A(B)=1 \textrm{ if and only if } A\subseteq B\eeq This is
similar to the case at classical physics, where the duals were
single element subsets (corresponding to characteristic maps). For
convenience we will occasionally identify the co-events with its
dual.
\end{itemize}

\item We require that every subset of measure zero is always mapped
to zero (i.e. it is always false).

\beq \label{exact preclusion}\mu(A)=0\Rightarrow \phi(A)=0\eeq
Co-events obeying equaiton \ref{exact preclusion} are called \emph{preclusive}. This requirement is same as
in classical physics, where the histories that are (or belong to a
set) of measure zero cannot occur, in other words are not
possible realities.

\item We require that the co-event (or actually its dual) is
\emph{primitive}. This means the following: We (partially) order all
possible duals (which are simply subsets of $\Omega$) with
respect to set inclusion. Then from all the (multiplicative)
preclusive co-events we choose only those that are minimal with
respect to this order, i.e. the finest grained (smaller) duals.
Remember  that in classical physics, these are simply single element
subsets ($h_i$), so by requiring primitivity we come as close as
possible to the picture we have in classical physics.
\end{enumerate}
We therefore get to the point where we identify these co-events,
which are (multiplicative), preclusive and primitive, as possible
realities (PPC). The multiplicativity, allows us to view the
relevant duals as what actually happens, and thus we have a very
similar picture with classical physics, where the difference is that
reality is no longer a fine-grained description but rather a
coarser-grained one\footnote{We can view this, as either finer
grained questions are un-physical, or physics at finer grained
scales has some (classically) contradictory properties.}. A very
interesting thing to point out here, is that the logic that arises,
depends on the dynamics (and initial condition), via the use of the
quantum measure (the zero sets) and it is \emph{not} fixed a-priori.

Before moving to the main text of this contribution, we shall stress
one more interesting point. Classical physics corresponds to
homomorphisms while the ``quantum'' nature is encoded in the parts
of the map that are anhomomorphic. Thus we could say that a
classical domain arises if we consider some coarse graining (i.e.
some subalgebra of $\mathcal{U}$) such that the induced map on this
subalgebra, is a homomorphism for ANY of the allowed (PPC)
co-events. What is most interesting, is that with this notion of a
classical domain there exists a unique finest grained classical
domain that all the others arise as coarse-grainings. This is in
striking difference with consistent histories where the main problem
was the existence of many incompatible classical domains. The reader
is once again refered to Ref. \cite{Emerent Probabilities} for
details.

\section{Probabilities in Anhomomorphic Logic}

\subsection{Probabilities in Classical Closed Systems: Cournot Principle}

In the analysis we have done so far, we have argued about what a
possible reality is (a PPC co-event), but have omitted what is
probable. In most everyday life cases, the predictions we get
(particularly from quantum
 theory) are probabilistic. However here we shall recall that
in consistent histories we have a single closed system, and in such
cases the very concept of probability (in classical physics) is not
easily and unambiguously defined. For example a probabilistic
statement about the state of the full universe cannot be testable,
since either outcome (finding universe possessing the property in
question or not) would not falsify our initial assertion.

There was a big philosophical debate by the founders of probability
theory on how one is expect to understand a probabilistic
statement\footnote{See for example Kolmogorov in the
``Grundbegriffe''.}. The standpoint which we shall adopt (which
suits best the case of a single closed system such as the universe)
is the use of the \emph{Cournot} Principle, and is closely related
with experimentally falsifying a theory :

\begin{quote}
\textbf{(Strong) Cournot Principle}: In a repeated trial, an event
$A$ of small measure ($\epsilon$ arbitrarily chosen), does
\emph{not} occur.
\end{quote}
Note that in a repeated trial, distributions of outcomes that differ
drastically from the probability distribution of the single trial,
would have very small measure and thus they do not occur. However,
this picture is problematic in classical physics. To see this, we
can consider the case of tossing a fair coin. The probability
measure of having all heads in $N$ trials, goes as $1/2^N$, while
the probability of getting heads less than (say) 60\% is
$\sum_{k=1}^{ 0.6N} {N\choose k}$ which is much greater than
$1/2^N$. We could thus claim that an outcome of all heads is
impossible, while an outcome with 50\% heads is possible. However,
we note, that in any actual realization (series of outcomes), we
will get a sequence of results (e.g. $httthtthht\cdots$) that has
probability of occurring $1/2^N$ exactly the same with the one we
get with only heads. This probability was very small ($<\epsilon$)
and thus this outcome was prohibited. In this example, we note that
\emph{any} possible outcome has  measure less than $\epsilon$ and if
we take the (strong) Cournot seriously that would come to a direct
contradiction with reality (since something actually happens).
Such considerations lead us to a milder version of the Cournot principle:

\begin{quote}
\textbf{Weak Cournot Principle}: In a repeated trial, an event $A$
\emph{singled out in advance}, of small measure ($\epsilon$
arbitrarily chosen), will \emph{not} occur.
\end{quote}
This means that if we ask in advance: ``Is the outcome 50\% heads
possible'' we will get the answer yes. However, if we ask in advance
``will the sequence $hththththt$ occur'' we will get the answer no,
since the measure for this outcome is small. We should stress here
that there is a split between the ontology and predictability of the
theory. According to weak Cournot, any outcome that doesn't have
identically zero measure, is possible. However if we have a theory
and we ask a question that has small ($<\epsilon$) non-zero measure,
if this outcome occurs, we falsify our theory, even if our theory
did not exclude this event from occuring. For example, if $N$
coin tosses give $N$ heads, we falsify our assertion that the coin
was fair, even if it is conceivable that it was a fair coin. To sum
up, in strong Cournot, no approximately zero measure set is a
possible reality, while in weak Cournot, approximately zero measure
sets are possible realities. In order to make scientific
predictions, we need to preselect  questions and we may need to
falsify our theory even if the experimental outcome is
(ontologically speaking) a possible reality.

Coming back to the case of a closed \emph{single} system, note that
while we cannot make a probabilistic statement for the universe
itself, we can instead make probabilistic predictions for subsystems
that `look identical'. In the above example we made statements about
the distribution of outcomes of tosses of coins (subsystems). These
statements, if we view them as statements about coins, correspond to
non-trivial probabilities. However, they correspond to `almost trivial' (i.e.
almost zero or almost one) probabilities for the full system of N
coins. The reader is refered to Ref. \cite{Emerent Probabilities} for further
details.

\subsection{Quantum Theory}

One of the things that stopped us from taking the strong Cournot
seriously, was the fact that we had an example that all $\Omega$ was
covered by histories (possible outcomes, $h_i$'s\footnote{Note that
by $h_i$ here we mean a sequence of single outcomes, as for example
$hththththt$ in the coin case.}) each of which was precluded because
it had measure $\mu(h_i)<\epsilon$. This resembles the Kochen
Specker theorem. Anhomomorphic logic, by considering as possible
realities subsets of $\Omega$ other than singletons (fine-grained
histories $h_i$), had evaded this contradiction. One would hope that
it may be possible to take strong Cournot seriously in the quantum
case, by defining Approximate PPC (APPC) co-events by replacing the
condition `$\mu(A)=0$' with the condition `$\mu(A)<\epsilon$' in the
definition of preclusion. (c.f. with Eq.\ref{exact preclusion}). We
hope that by making this change we can encode all the `useful'
information in the quantum measure in our set of APPC co-events;
then the quantum measure itself would no longer be needed for
predictions. However this leads to contradictions with observation.
For if the dual of a co-event contains two histories giving
different answers to a particular question, then the co-event itself
will answer our question in a manner inconsistent with Boolean
logic; even when this question relates to an experimentlaly
observable outcome (see Ref. \cite{Emerent Probabilities} for
further details). For example, returning to our repeated trial of a
coin, assume we have a APPC co-event that has in its dual one
history that has a heads outcome in the first coin toss, and a
second history that has a tails outcome. Then by the definition of
anhomomorphic logic, we would get NO to the question ``did the first
coin toss result in heads?'' and also NO  to the question ``did the
first coin toss result in tails''. But we can experimentally verify
that exactly one of the two outcomes `heads', `tails' will actually
occur if we are to throw our coin. Such co-events (that are possible
for APPC co-events), would not be possible had we stick with
\emph{exactly} preclusive co-events (see Ref. \cite{Emerent
Probabilities} for further details). We are then pushed back to the
use of the weak Cournot.

In quantum theory when we use weak Cournot, the possible realities are still the PPC co-events (c.f. in classical physics, possible
realities were fine grained histories, that had non-zero measure,
even if it was arbitrarily small). To deduce probabilistic
predictions we need to use once again the quantum measure. We select
a question corresponding to a subset $A$ of the history space
$\Omega$. If $\mu(A)<\epsilon$, we say that this result is not
possible, and this is the prediction we make. If this outcome
actually arises, we then say that our initial assumption about what
was the quantum measure (which includes initial conditions and
dynamics) has been falsified.

Note, the following difference of the quantum case compared to the
classical one. In classical physics the measure that is used to
derive predictions, is on the space of possible realities, which is
the history space. In quantum theory, the quantum measure is again
on the history space, however this is no longer the space of
possible realities, since the possible realities are the PPC
co-events. The reader is refered to Ref. \cite{Emerent
Probabilities} for further details.

\subsection{The Double Slit Example}

Finally, let us show how the quantum measure can be used along with weak Cournot to get predictions that reproduce the double slit
pattern. For simplicity we will consider a discrete screen, with 5
slots ($i\in\{0,\pm1,\pm2\}$)\footnote{Slots 0 and $\pm2$ are the
bright fringes and $\pm1$ the dark fringes of the pattern.}. We also
have two slits $(s_1,s_2)$. Each fine grained history (for a single
repetition), consists from the particle crossing one slit and
hitting the screen at on slot. The measure is given:

\bea
|(s_j,i)|&=&0.1\nonumber\\|(s_1,i)|\sqcup(s_2,i)|&=&\left\{\begin{array}{ll}
0.3 & \textrm{if $i=\pm2,0$}\\ 0.05 & \textrm{$i=\pm 1$}
\end{array} \right.\nonumber\\ |(s_k,i)\sqcup(s_l,j)|&=& |(s_k,i)|+|(s_l,j)|\nonumber\\ &=& \textrm{$0.2$ if $i\neq j$}\nonumber
\eea

If we have many independent repetitions we can simply use the product measure. Now assume that we have a closed system
with 10 particles that are to cross this double slit, and we assume
that anything that has measure less than $\epsilon=10^{-3}$ is
precluded, meaning that if such outcome comes we falsify our initial
theory (according to the weak Cournot). We can easily see that the
question ``Is the distribution uniform'' consists of all the
permutations where we have 2 particles at each slot, and this number
is ${10\choose2} {8\choose2} {6\choose2} {4\choose2}=113400$ which
is much larger than the number of permutations for getting the
double slit pattern (i.e. 3 particles at $\pm2,0$ slots and 1 at
either 1 or -1). The latter is 4800. However the measure for each
fine grained history of 10 particles that are uniformly distributed
is only $\approx5\times 10^{-9}$. The total measure of all histories
that are of uniform distribution is then
$113400\times5\times10^{-9}\approx5\times10^{-4}$. This is less than
$\epsilon$, so the uniform distribution is precluded. In contrast while there
are much fewer combinations that give the double slit distribution, the measure
for each of them is $\approx10^{-6}$ and thus the total measure for
getting a distribution `like' the double slit pattern is
$4800\times10^{-6}\approx5\times10^{-3}$, which is greater than
$\epsilon$ and thus the double slit distribution is not ruled out by our
theory\footnote{Note, however, that if we asked more detailed
question, such as ``the first particle will hit slot 0, the second
slot 2, etc'' even if the distribution was correct the measure would
be very small ($\approx10^{-6}$ here) and thus ruled out by our
predictions. The ``paradox'' here, as in classical physics, would be
that the eventual outcome of the experiment, would be one of those
ruled out histories, and the contradiction is evaded by the crucial
concept of \emph{pre}-selected questions, in the definition of weak
Cournot.}. Here, in this simplified model, we see how predictions of
the type of double slit pattern (i.e. ruling out distributions such
as the uniform one) can arise in anhomomorphic logic, when we use
the quantum measure of  many repetitions of the system along with
the weak Cournot principle.

\section{Summary and Conclusions}

In this contribution, we reviewed the anhomomorphic logic approach to
quantum theory, which is a development of the consistent histories
approach. Reality is no longer a single fine grained history, but a
primitive preclusive multiplicative co-event, which can also be
viewed as a coarse grained history. The Kochen Specker theorem is
evaded, as is the problem of many incompatible
classical domains faced by consistent histories.
The core of this contribution was how to deal with probabilistic
predictions in this formalism. We resorted to the Cournot principle
to give meaning to probabilistic statements and essentially in a
frequentist's view on probability (rather than propensity). The use
of strong Cournot principle was ruled out (alas for different
reasons than in classical physics). The weak Cournot, introduces a
split of ontology and predictions. In classical physics this leads
to:

\begin{itemize}
\item [(a)] Ontology: Fine grained histories are the possible
realities. In other words one and only one history is actually
realized.
\item[(b)]Predictions: The (classical) measure on $\Omega$ is used,
in order to make predictions, that are of the type ``if $A$ is
realized and $\mu_c(A)\leq\epsilon$ then our initial assumption is
rejected''.
\end{itemize}

In quantum theory the picture is similar:

\begin{itemize}
\item [(a)] Ontology: Possible realities are the multiplicative,
primitive and (exactly) preclusive co-events. If we view the dual
picture (the duals), we can say that what is realized is a coarse
grained set of histories or else a non-trivial subset of $\Omega$.
\item[(b)]Predictions: The \emph{quantum} measure on $\Omega$ is used,
in order to make predictions, that are of the type ``if $A$ is
realized and $\mu(A)\leq\epsilon$ then our initial assumption is
rejected''. Note however, the fact that the quantum measure is on
the history space, which in the quantum case, is no longer the space
of possible realities.
\end{itemize}

\ack The authors would like to thank Fay Dowker, Rafael Sorkin and
Sumati Surya for many discussions on Anhomomorphic Logic. PW thanks
the organizers for carrying out this very interesting conference and
giving him the opportunity to give this talk. Partial support from
the Royal Society International Joint Project 2006-R2 is
acknowledged and YGT was supported by an STFC studentship.

\end{document}